\documentstyle[epsf,psfig]{aipproc}

\newcommand{\be}{\begin{equation}}
\newcommand{\ee}{\end{equation}}

\begin{document}

\title{Analysis of Vocal Disorders \\ in a Feature Space}

\author{Lorenzo Matassini, Rainer Hegger, Holger Kantz}

\address{Max-Planck-Institut f\"ur Physik komplexer Systeme\\
         N\"othnitzer Str.\ 38, D 01187 Dresden, Germany\\
         email: lorenzo@mpipks-dresden.mpg.de}

\author{Claudia Manfredi}

\address{Dept. of Electronic Engineering - Univ. of Florence\\
         via Santa Marta 3, I 50139 Firenze, Italy\\
	 e-mail: manfredi@die.unifi.it}

\maketitle

\begin{abstract}

This paper provides a way to classify vocal disorders for clinical applications. This goal is achieved by means
of geometric signal separation in a feature space. Typical quantities from chaos theory (like entropy, correlation
dimension and first lyapunov exponent) and some conventional ones (like autocorrelation and spectral factor)
are analysed and evaluated, in order to provide entries for the feature vectors.
A way of quantifying the amount of disorder is proposed by means of an \emph{healthy index} that measures the
distance of a voice sample from the centre of mass of both healthy and sick clusters in the feature space.
A successful application of the geometrical signal separation is reported, concerning
distinction between normal and disordered phonation. 

\end{abstract}

\section*{Introduction}

The voicing source can produce in addition to periodic vibrations (the normal phonation) a great variety 
of complex signals, due to the fact that many sources of non-linearity are involved
in the air-flow production and in the laryngeal vibration processes \cite{berg}.
These features have been modelled successfully in the past years
by an asymmetric two-mass model of the vocal folds \cite{ishizaka}. Standard methods 
of voice analysis, such as the estimation of jitter and shimmer and the harmonics-to-noise 
ratio, are valuable for the characterization of regular phonation. However, in
the case of voice disorders abrupt changes to irregular regimes
occur and these measurements are of limited relevance \cite{kelman}.

Many studies \cite{berg,herzel2} suggest that some of the complexities observed in rough, 
disordered voices are not caused by external sources, but by 
the intrinsic nonlinear dynamics of vocal fold movement. Given the nonlinearities associated 
with the laryngeal source (i.e. the pressure-flow relation in the glottis, the stress-strain curves
of vocal fold tissues, the vocal fold collisions), this should not come as a great surprise.

The transitions to qualitatively new oscillatory behavior indicate the suitability of the 
methods from nonlinear dynamics; e.g. in \cite{herzel} it is shown that a sufficiently large 
tension imbalance of the left and right vocal fold induces bifurcations
to chaos. Asymmetries due to paralysis, polyps, papilloma, cancer etc. produce the same effects 
and simulations of coupled oscillators \cite{reuter} can provide insight into the sources 
of such vocal instabilities and are, therefore, of potential use for diagnosis and treatment 
of voice disorders. Furthermore the reconstruction of attractors and the estimation of their 
properties indicate low dimensionality in the systems generating the voice signals.

Herzel et al. \cite{herzel2} have shown that indications of secondary Hopf bifurcations can be 
found in pathological voices, as well as sudden jump from one limit cycle to another one with 
different period and amplitude. Different attractors may coexist in nonlinear systems and, therefore, 
even extremely tiny changes of parameters, like muscle tension, may lead to abrupt jumps to
other regimes. The occurrence of such a situation should be reflected in quantities like the 
entropy and the fractal dimension of the (global) attractor.

Qualitatively, the origin of bifurcations and low-dimensional attractors can be understood as 
follows: Normal phonation corresponds to an essentially synchronized motion of all vibratory modes. 
A change of parameters such as muscle tension or localised vocal fold lesions may lead to a 
desynchronization of certain modes resulting in bifurcations and chaos. The following modes are 
of particular relevance: Motion of the left and right vocal fold, horizontal and vertical modes,
interaction of the ventricular and vocal folds, interaction of vocal fold vibrations with sub- and 
supra-glottal acoustic resonances, and vortices generated at the glottis.

Although normal phonation and voice disorders can be distinguished qualitatively by human very easily, 
a quantification and data distribution of the disease is highly desirable.
This work is motivated by clinical interests, which lie in objectively evaluating the effort made 
by cordectomised patients during an utterance, as it could be indicative of patient status, also as 
far as post-operatoty functional recovery is concerned \cite{kasuya,manfredi}.

\section*{Geometric signal separation}

The idea of geometric signal separation in a feature space is used in different applications, to 
identify the dynamical state of a complicated system, where for practical reasons the system itself 
is connected to a simple measurement device which records a scalar time series. Sufficiently long 
subsections of this series are transformed into feature vectors $v$ in a feature space $V$. The 
entries of $v$ are chosen to be quantities which contain the compressed information of the signal 
relevant for the task to be performed and which can be estimated directly from the time series. 
Neighbourhood relations in this space, based on the computation of distances between feature vectors, 
allow for various classification and diagnosis tasks.
It can be necessary to introduce a local metric, since the variability of quantities may range over very 
different scales.

During a training period  pre-classified (e.g. by human experts) data sets are collected
and each one is converted into a feature vector. This set of vectors is divided into clusters, 
where each cluster represents a dynamical state of the system.
In the data classification period feature vectors are calculated and compared to the clusters in $V$. 
The distance of each test vector to the closest cluster is thresholded to yield a distinction between 
classes. A successful application of the method for on-line failure detection on electro-motors is 
discussed in \cite{guttler}.

The basic idea of using a feature space for our purpose is therefore to eliminate the short-time 
variability of the time series, extracting characteristic features.
As a first step for the classification of vocal disorders, one has to select suitable entries of 
the feature vectors $v$. These have to contain extremely condensed information from the time series, 
reflecting somehow a pseudo-state of the dynamical system that has produced the sentence. In other 
words, healthy patients and sick ones should be associated to feature vectors that populate different 
regions of the feature space.

In \cite{berry}, it was shown by means of the empirical orthogonal functions that normal phonation 
is well represented by only two eigenmodes, while the simulation of disordered voice has shown that 
the three strongest modes can cope with 90\% of the variance. So the fractal dimension of the attractor 
is a good entry of the feature space, since healthy people should produce smaller dimension-values than 
patients with some kind of disease. 
There are several ways to quantify the self-similarity of a geometrical object by a dimension. From a 
computational point of view it is convenient to proceed in the way illustrated in \cite{kantz1}. Let us 
define the correlation sum for a collection of points $x_n$ in some vector space to be the fraction of 
all possible pairs of points which are closer than a given distance 
$\epsilon$ in a particular norm. The basic formula is:

\be \label{eq:corr_sum}
C(\epsilon)=\frac{2}{N(N-1)}\sum^{N}_{i=1}\sum^{N}_{j=i+1}\Theta(\epsilon - ||x_i-x_j||),
\ee

where $\Theta$ is the Heaviside step function. The sum just counts the pairs $(x_i,x_j)$ whose distance 
is smaller than $\epsilon$. In the limit of an infinite amount of data $(N\rightarrow\infty)$ and for 
small $\epsilon$, we expect $C$ to scale like a power law, and we can define the correlation dimension $D$ by:

\be \label{eq:corr_dim}
D=\lim_{\epsilon\rightarrow0}\lim_{N\rightarrow\infty}\frac{\partial\ln C(\epsilon,N)}{\partial\ln\epsilon}. 
\ee

Since in the latter there are two limits involved, and both limits are not computable in a closed form, 
one has to look very carefully to the results before claiming some numbers as the value of the correlation 
dimension. It should be clear that one needs a lot of points in order to estimate $C(\epsilon)$ over a 
large enough range of length scales. A few hundred are definitely not enough to yield a statistically 
significant result at the small length scales, where also the noise starts to play a noteworthy
role. With our time series a clear identification of a dimension was not always possible; this is 
essentially the reason why we will introduce a pseudo-dimension or dimension-like quantity. It is 
important to remember a result reported in \cite{kantz1}. In theory, the maximum dimension $D_M$ that can 
be calculated for a data file of length $N$ is:

\be \label{eq:max_dim}
D_{M} \approx 2 \lg_{10}(N-2).
\ee

Therefore, approximately eight dimensions is the maximum which can be calculated from a 22 KHz sampled 
sentence lasting one second.

One other candidate for the feature space is represented by the entropy, a fundamental concept in statistical
mechanics and thermodynamics. Entropy describes the amount of disorder in the system, but one can
generalise this concept to characterise the amount of information stored in more general probability distributions. 
Let us introduce a partition $\mathcal{P}_\epsilon$ on the dynamical range of the observable, and the 
joint probability $p_{i_1,i_2,...,i_m}$ that at an arbitrary time $n$ the observable falls
into the interval $I_{i_1}$, at time $(n+1)$ it falls into interval $I_{i_2}$ and so on. Then one 
defines block entropies of block size $m$ and partition radius $\epsilon$ the quantity:

\be
H_q(m,\epsilon)=\frac{1}{1-q}\ln\sum_{i_1,i_2,...,i_m} p_{i_1,i_2,...,i_m}^q.
\ee

The order-$q$ entropies are then:

\be \label{eq:entr}
h_q=\sup_{\mathcal{P}_\epsilon}\lim_{m\rightarrow\infty}\frac{1}{m}H_q(m,\epsilon) = 
	\sup_{\mathcal{P}_\epsilon}\lim_{m\rightarrow\infty} (H_q(m+1,\epsilon)-H_q(m,\epsilon)).
\ee

In the strict sense only $h_1$ is called the Kolmogorov-Sinai entropy \cite{kolmo,sinai}, but in fact all 
order-$q$ entropies computed on the joint probabilities are entropies in the spirit of Kolmogorov and Sinai, 
who were the first to consider correlations in time in information theory.
Due to the numerical problems encountered in the estimation of the entropy from real data, namely the finite 
length of the time series and the presence of noise, we will use a pseudo-entropy in the following.

The full feature vector contains the following quantities:

\begin{itemize}

\item {\bf Spectral Factor:} It is the averaged ratio between the amplitude of frequencies under $1$ KHz and 
frequencies between $4$ and $6$ KHz; it is motivated by the effort sick subjects have to face when they want 
to speak; this induces instabilities that are reflected by the power spectrum (sick subjects present a smaller
value).

\item {\bf Pseudo-Entropy:} The quantity $h_2$, as defined in Eq. \ref{eq:entr}, is averaged for 
$\epsilon$-values of 5\% to 10\% of the variance of the data for embedding dimensions ranging between 
2 and 8, upper limit suggested from Eq. \ref{eq:max_dim}. Sick subjects should present a bigger value
than healthy people.

\item {\bf Pseudo-Correlation Dimension:} The quantity $D$, as defined in Eq. \ref{eq:corr_dim}, is averaged 
for $\epsilon$-value of 5\% to 10\% of the variance of data for embedding dimensions ranging between 2 and 8.
Sick subjects should present a bigger value than healthy people.

\item {\bf First zero-crossing of the Autocorrelation Function:} This parameter is related to the ability of 
the subject in correctly pronouncing a word. In particular, dysphonic patients are not able to isolate every 
vowel, and the resulting time series is more correlated than for healthy subjects. The estimation method is 
the same as in \cite{kantz1}. 

\item {\bf First Lyapunov Exponent:} This is a convenient indicator of the sensitivity to small orbit 
perturbations characteristic of chaotic attractors, as it gives the average exponential rate of divergence 
of infinitesimally nearby initial conditions. Some sicknesses can induce sudden jumps from the limit cycle 
(to which a zero maximum Lyapunov exponent is associated) to another one with different period and amplitude 
(but again with zero maximum Lyapunov exponent); if the jump is due to a bifurcation one can see a positive 
value. Estimating this quantity, anyway, one has to be careful because Lyapunov exponents for speech data 
are inconsistent\footnote{Kumar and Mullik \cite{kumar} found the maximum exponent to be positive, 
characteristic of chaos, for normal vowel and consonant productions. In contrast, Narayanan
and Alwan \cite{nara} calculated a maximum exponent of zero, indicating a limit cycle, for normal vowel 
phonation and a positive value for voiced and voiceless fricative productions. Herzel \cite{herzel} found 
a maximum exponent of zero for healthy vowel phonation, while a dysphonic vowel sample yielded a positive 
value. All authors cautioned, however, that calculation of the exponent is highly sensitive to short data 
sets and nonstationarity. In fact, critical sensitivity to noise \cite{kantz2} makes calculation of a 
Lyapunov exponent highly suspect and inconclusive for voice data.}.

\item {\bf Prediction Error:} We apply the idea presented in \cite{farmer}; when a well defined attractor 
is present, then the prediction error has to be small. Normal phonations were found indeed to lie with a good
approximation on a limit cycle, while sick people commonly produce more disordered time series.

\item {\bf Jitter:} As in \cite{herzel}, to take into account the short-term (cycle-to-cycle) variation 
in the fundamental frequency of the signal. Commonly, for healthy voices, the jitter is lower than $1\%$, 
while higher values characterise disphonic phonation.

\item {\bf Shimmer:} As in \cite{herzel}, to take into account the short-term (cycle-to-cycle) variation 
in the amplitude of the signal. As for the jitter, large values commonly indicate great effort in speaking.

\item {\bf Peak in the Phoneme Transition:} The transition between one phoneme and the following shows 
a much longer transient for sick people (they have problems to switch from a dynamical regime to the following). 
Every phoneme contains a pitch that is repeated a number of time 
variable between 10 and 20, as presented in \cite{lm,lm2}. One has then
to identify the time length of such a pitch and to compare the distance between this pattern and one 
of the same length coming from the same phoneme. Moving the second pattern 
along the full phoneme, one gets a distribution of distances (with a zero when the reference pitch is 
considered and saturation values close to the end of the actual phoneme); the maximum peak of this 
distribution before the saturation, i.e. before the end of the current phoneme, gives what we call 
Phoneme Trasition. We have discovered that this quantity is comparable, for our purposes, to the already 
mentioned harmonics-to-noise ratio. We prefer to estimate the Peak in the Phoneme Transition because it 
looks more reliable.
Sick subjects present bigger values than healthy people.

\item {\bf Residual Noise:} The algorithm presented in \cite{lm} is applied to the time series. Noise can 
be easily removed if the series contains redundancy and, roughly speaking, redundancy is a synonym of 
healthy\footnote{In \cite{kasuya,manfredi} the Normalized Noise Energy is introduced to measure the 
dysphonic component of the voice spectrum related to the total signal energy. The deviation from periodicity 
in the sub-phoneme structures, due to the dysphonia, can be interpreted as noise and then a way to quantify 
this additive component is preposed.}. 
After applying the noise reduction algorithm, we look at the variance of the difference between the original
and the processed signal. Healthy subjects are related with small values, sentences spoken by sick people
look very noisy and therefore present a bigger value of the residual noise after the filtering.

\end{itemize}

Of course the feature vector is redundant, but this is something somehow unavoidable, since all the entries 
have to come from the same time series and in this meaning they must be more or less correlated. Furthermore 
working in a larger dimensional feature space helps to better identify the pathology, since the distances 
between points become larger. Of course a \emph{Principal Component Analysis} is very helpful in detecting
the trade off between redundancy and robustness.

\section*{Experimental results}

We now proceed to collect male voice samples from three categories of subjects\footnote{These voice samples were 
recorded in a quiet room at the Phoniatric Section of the Otholaryngoiatric Institute, Careggi Hospital, Firenze.}:

\begin{itemize}

\item People suffering from dysphonia\footnote{Adult subjects affected by T1A glottis cancer, a tumour 
confined to the glottis region with mobility of the vocal cords.} (12).

\item Healthy people (17).

\item People with pathologies under medical treatment\footnote{Subjects operated via endoscopic laser or 
traditional lancet technique.} (4).

\end{itemize}

All these subjects have been asked to say the italian word {\bf \it aiuole} (flower-beds), as it is made up 
with the five main Italian vowel sounds: `a', `e', `i', `o', `u'. Every time the word was uttered in isolation.
The use of a complete word instead of 
sustained vowels is due to the clinical interest in evaluating the effort made by the patient during the 
entire vocal emission, also as far as the glide between vowels (`ai', `iu', `uo') is concerned.
The sentences have been recorded in a {\bf .wav} mono, 16-bits linear, 22050 Hz format 
converted into sequences of real numbers\footnote{Every voice sample is slightly shorter than 1s; this means 
that the corresponding time series contains about 20000 points}
and given to the feature vector building algorithm, which always treats them as whole words.

\begin{figure}
\centerline{\psfig{file=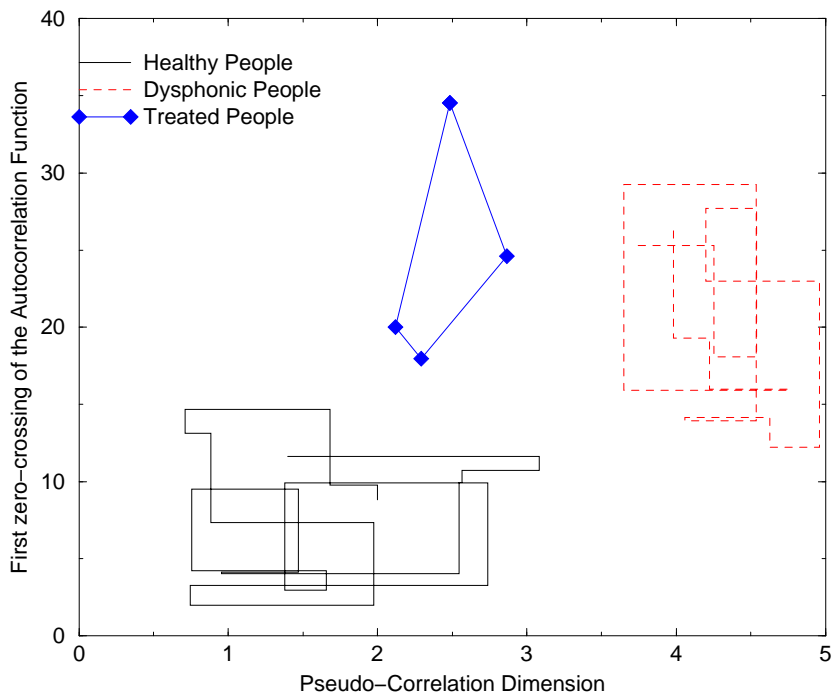,width=11cm,angle=270}}
\caption[]{\small\label{fig:feat4}
Two-dimensional projection of the feature space onto the third and fourth components of the feature vector.}  
\end{figure}

A two-dimensional projection of the feature space is visible in Fig.\ref{fig:feat4}, namely onto the 
pseudo-correlation dimension and the first zero-crossing of the autocorrelation function. A principal component
analysis has revealed it to be one of the clearest bidimensional projection. Others give as well good
separation, but since they have combination of features as axes we prefer to present the results as in 
Fig.\ref{fig:feat4}.

As expected, a normal voice does not exceed the third dimension, while up to 5 degrees of freedom are necessary 
to represent a pathologic phonation. A similar sharp distinction is given by the autocorrelation function: 
Dysphonic subjects are not able to well isolate every vowel and the
resulting time series is more correlated, i.e. the first zero-crossing is reached after 1.2 ms 
(typical values for healthy subjects range between 0.2 ms and 0.4 ms). 

Similar results are readable through the entropy as far as sick people produce more disordered series; 
the prediction error is smaller for normal phonations because in the phase space they lie with a good 
approximation in a limit cycle. Also the filtering of the time series is easier when no disease is there; 
the absence of bifurcations facilitates the search of neighbours and
improves the quality of the noise reduction (\cite{lm}).

In order to check the classification ability of the method,
we have collected two more sets of data:

\begin{itemize}

\item Healthy people simulating a disease (17).

\item Artificial voices (12).

\end{itemize}

\begin{figure}
\centerline{\psfig{file=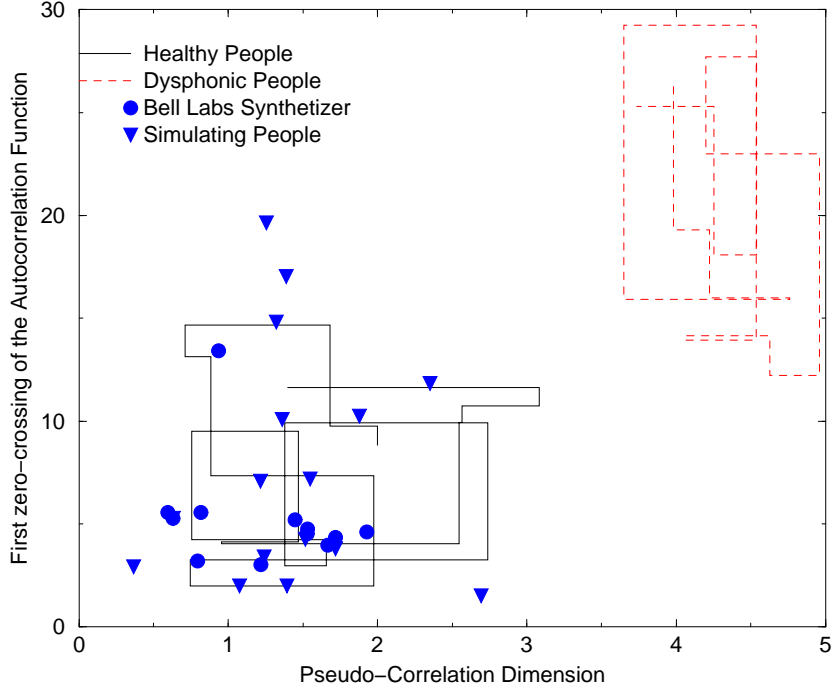,width=11cm,angle=270}}
\caption[]{\small\label{fig:feat3}
Two-dimensional projection of the feature space onto the third and fourth components of the feature vector.}  
\end{figure}

In the first set, people have tried to say {\bf \it aiuole} in the strangest possible way
(getting sometimes very impressive records!), simulating hoarseness. They could listen as many time as they
wanted to sentences spoken by diseased speakers and they were given the instruction to imitate them as close
as possible.
In the second we have used the speech synthetizer in \cite{bell_labs} with 
different languages and several voices (man, gnat, raspy, woman, coffee drinker, ridiculous, child, big man). 
Fig.\ref{fig:feat3} shows the results. It is very interesting to note that, although the simulated and the 
diseased sentences sound quite similar, they are correctly classified by the algorithm. 
In particular nobody was able to exceed the dimension three, since the dysphonia is something that one 
cannot directly control\footnote{In the two-mass model of the vocal cords, introduced in \cite{ishizaka}, 
the desynchronization is the source of irregularity and it is achieved through the introduction of asimmetry 
in the model itself. In the speech production, the left and right fold can be regarded as separate
oscillators which are coupled via the airflow and collisions. For strong asymmetry, like a unilateral vocal 
fold paralysis, both folds may desynchronize, leading to complex vibratory pattern. In simulating the disease 
it's not clear what one should do in order to get this asymmetry effect.}. Less amazingly, artificial voices 
lie inside the healthy zone, but they also sound somehow less rich and, so to say, artificial.

Fig.\ref{fig:feat4} shows also the results concerning the set of voices coming from cordectomised patients.
They lie between the healthy and the sick region, with small pseudo-correlation dimension values but large
first zero-crossing of the autocorrelation ones. This result, though preliminar, shows that cordectomised
subjects still require more effort in speaking with respect to healthy people, due to the limited extension
of the produced scar fold. More results could be obtained if the long-time path followed by patients after
treatment were available.

In order to quantify the degree of illness, we define the following index. 
First, the centre of mass of both the healthy ($\overline{healthy}$) and pathologic clusters ($\overline{sick}$)
is computed. Then the distances $d(new,\overline{healthy})$ and $d(new,\overline{sick})$ of the voice sample under 
test from $\overline{healthy}$ and $\overline{sick}$ respectively are evaluated.
Due to the different range covered by the
different components of the feature vector, a weighted distance is considered, where the weights are the inverse 
of the standard deviations of the distribution of the entries. The
{\bf healthy index $\mathcal{H}$} is defined as:

\be
{\mathcal{H}}(new)=20\log_{10}\frac{d(new,\overline{healthy})}{d(new,\overline{sick})}.
\ee

A strong negative value of $\mathcal{H}$ indicates a healthy voice, while a big positive one reveals the 
presence of some kind of pathology. Of course it is necessary to introduce some thresholds to get a good 
classification, according to clinical considerations.

\begin{figure}
\centerline{\psfig{file=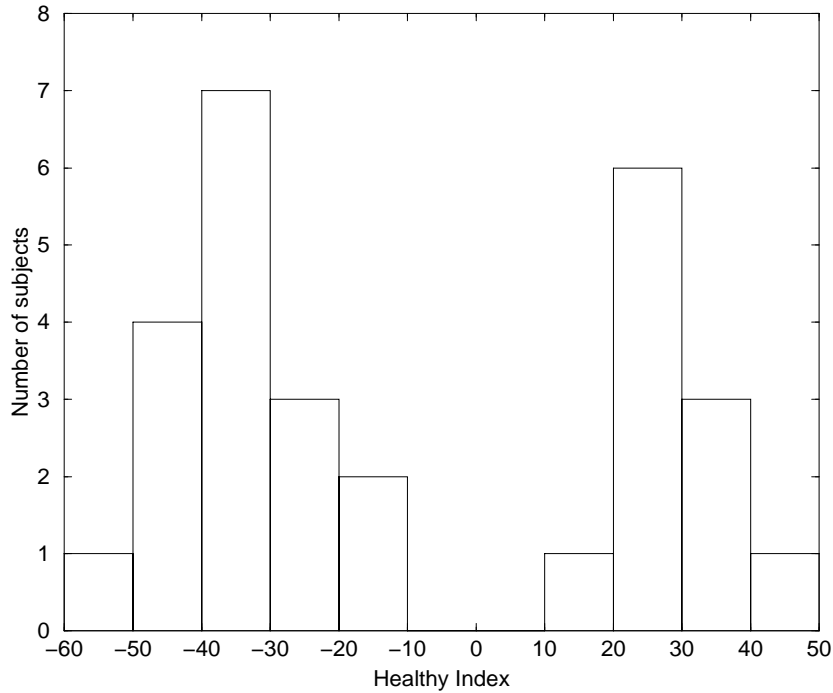,width=11cm,angle=270}}
\caption[]{\small\label{fig:distr}
Distribution of healthy and sick people according to the healthy index.}
\end{figure}

We have computed the healthy index $\mathcal{H}$ for the full set of data, getting the distribution
for healthy and sick people shown in Fig.\ref{fig:distr}. The peak on the left is relative to normal phonation, 
while the right one regards only disphonic voices. 
Cordectomised patients got an index value ranging between -7 and 4, artificially generated 
voices between -77 and -45, healthy people simulating a disease are situated
between -53 and -5. It is again interesting to note how difficult (if not impossible) it is for healthy people to
simulate a real disease, since it operates directly into the physiologic level. 

Notice that the healthy index is based on all the entries of the feature vector. Reducing the number of parameters 
causes a smoothed distribution of voices, with a worse classification. Since in this application the computation 
time was not a problem (the word \emph{aiuole} is vey short and the total number of samples is very limited), we
did not put to much effort in determining up to which extent we can neglect some parameter. As already discussed,
some of them play a more important role than some other and this aspect needs to be considered in more details if
the speed of the classification becomes a crucial point.

\section*{Conclusions}

The vocal folds, together with glottal airflow, constitute a highly nonlinear self-oscillating system. 
According to the accepted myoelastic theory of voice production, the vocal folds are set into vibration 
by the combined effect of subglottal pressure, the viscoelastic properties of the folds, and the Bernoulli 
effect \cite{berg}. The effective length, mass, and tension of the vocal folds are determined by muscle 
action, and in this way the fundamental frequency and the waveform of the glottal pulses can be
controlled. The vocal tract acts as a filter which transforms the primary signals into meaningful voiced speech.
Normal sustained voiced sound appears to be nearly periodic, although small perturbations are important 
for the naturalness of speech. On the other hand, vocal instabilities due to bifurcations are often 
symptomatic in voice pathology.

This work aims at classifying vocal disorders through the construction of an
appropriate feature space. The main point is the identification of the entries of the feature vectors, 
as these should be quantities into which differences between normal and disordered phonation are well visible. 
This choice is made here following physician's indications, according to physical models simulating the vocal
folds, linear and non-linear dynamic theory.
Some results concerning two-dimensional projections of the feature space are presented,
showing a clear separation between healthy and sick voices with the selected parameters. The robustness of the
method is tested through simulations of dysphonic voices and the use of a speech synthetiser.

A way of quantifying the amount of disorder is proposed by means of an healthy index that measures the distance of a
voice sample from the centre of mass of both healthy and sick clusters in the feature space. The results obtained
show the good performance of the method.
The approach can be of clinical interest also as far as the post-operatory evolution (and possible rehabilitation)
is concerned, commonly performed on a subjective basis only. Glottal functionality can in fact be analysed by 
means of objective indexes other than visual inspection of the spectrogram.

\vspace*{6mm}
The authors would like to thank the Careggi Hospital in Florence, in particular the Phoniatric Section and 
Prof. Cecilia Salimbeni, for fruitful collaborations.

\end{document}